\documentclass[a4paper,fleqn,usenatbib]{mnras}

\usepackage{newtxtext,newtxmath}

\usepackage[T1]{fontenc}
\usepackage{ae,aecompl}
\usepackage{graphicx}	
\usepackage{amsmath}	
\usepackage{amssymb}	
\newcommand{\msol}{~\mathrm{M}_\odot}
\newcommand{\rsol}{~\mathrm{R}_\odot}
\newcommand{\teff}{T_{\mathrm{eff}}}
\newcommand{\llsol}{L/\mathrm{L}_\odot}

\title[]{On  Binary Channels to Anomalous Cepheids}

\author[A. Gautschy and H. Saio]{
Alfred Gautschy,$^{1}$\thanks{E-mail:alfred@gautschy.ch}
and Hideyuki Saio$^{2}$
\\
$^{1}$CBmA, Liestal, Switzerland\\
$^{2}$Astronomical Institute, Graduate School of Science, Tohoku University, Sendai, Japan\\
}

\date{Accepted 2017 March 30. Received 2017 March 30; 
      in original form 2017 January 6}

\pubyear{2017}

\begin{document}
\label{firstpage}
\pagerange{\pageref{firstpage}--\pageref{lastpage}}
\maketitle

\begin{abstract}
  Anomalous Cepheids are a rather rare family of pulsating variables
  preferably found in dwarf galaxies. Attempts to model these variable
  stars via single-star evolution scenarios still leave space for
  improvements to better grasp their origin.  Focusing on the LMC with
  its rich population of Anomalous Cepheids to compare against we
  probe the r\^ole binary stars might play to understand the nature of
  Anomalous Cepheids. The evolution of donors and accretors undergoing
  Case-B mass transfer along the first red-giant branch as well as
  merger-like models were calculated. First results show that in
  binary scenarios a larger range of star masses and metallicities up
  to~$Z\la 0.008$, higher than deemed possible hitherto, enter and
  pass through the instability strip. If binary stars play a r\^ole in
  Anomalous Cepheid populations, mass donors, mass accretors, or even
  mergers are potential candidates to counteract constraints imposed
  by the single-star approach.
\end{abstract}


\section{Introduction}

\begin{figure}
	\includegraphics[width=\columnwidth]{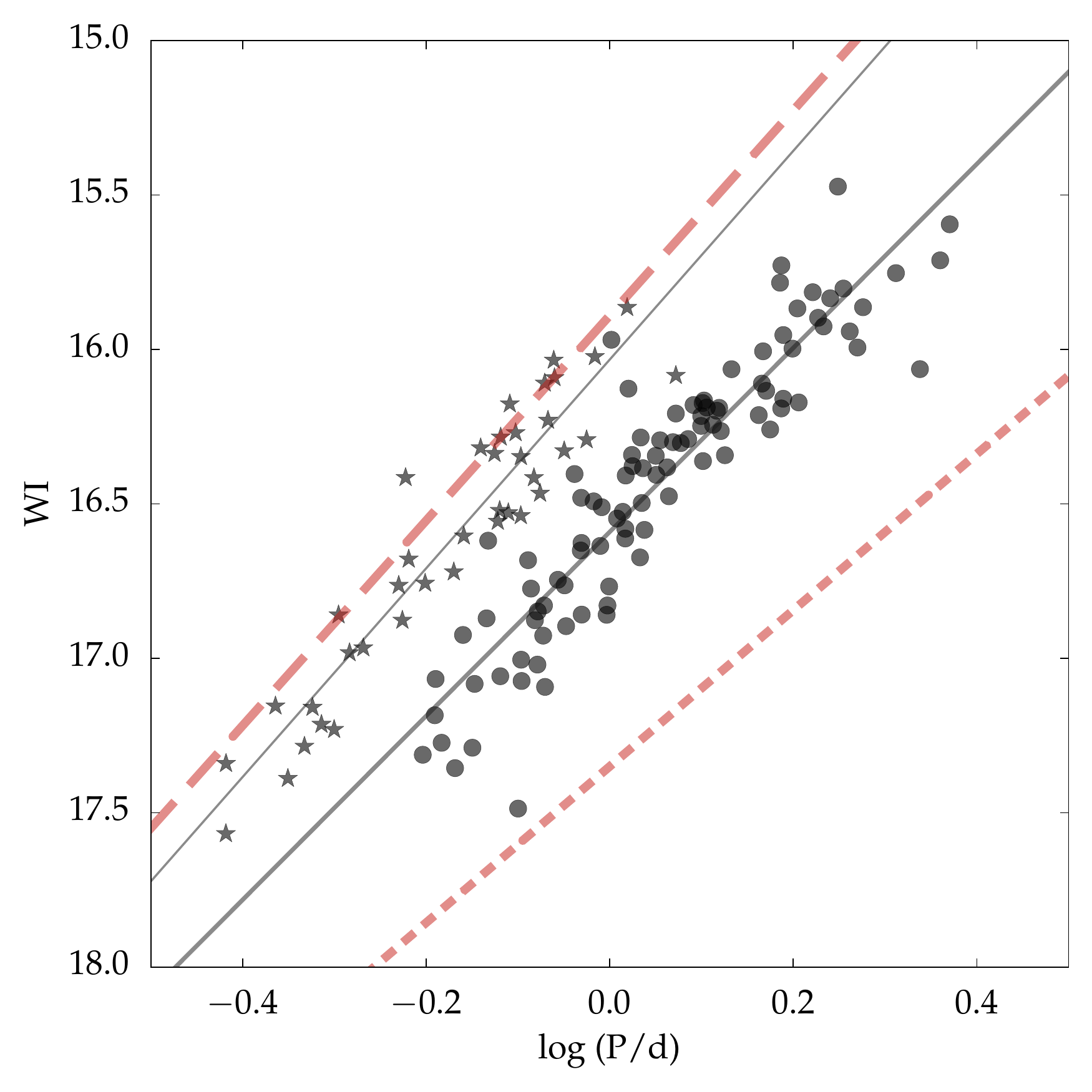}
    \caption{ACs in the LMC as observed and calibrated by the OGLE team
    	     \citep{Soszynski2015} on the period~--~Wesenheit-index plane.
    	     The quantity WI is computed as $I - 1.55\left( V - I \right)$ 
    	     referring to the magnitudes in the photometric passbands $V$ and $I$.  
    	     Full circles display fundamental-mode pulsators, asterisks mark 
    	     first overtone ones. The full lines trace the  
    	     linear fits to the observations. The computed fit 
    	     parameters are listed in Table \ref{tab:PWI-Fits} labeled 
    	     with OGLE LMC. The dashed lines put ACs in relation to 
    	     the fits to the P-WI relations of Classical (long dashes) 
    	     and TypeII F-mode Cepheids (short dashes) in the LMC.
    	    }
    \label{fig:ACs_LMC}
\end{figure}

Short-period Cepheid-like variable stars, at the time almost
exclusively found in dwarf spheroidal galaxies, were tagged with the
diagnostic adjective ``anomalous" if they did not fall into any of the
customary regions occupied by classical pulsators (PopI~--~,
Type~II~--~Cepheids, and even RR Lyrae stars) on the
period~--~luminosity (PL) plane \citep{Zinn1976}. Also the lightcurve
morphologies of Anomalous Cepheids (ACs), quantified via Fourier
coefficients, suggest that many of them are distinct from their
classical siblings \citep[e.g.][on the Magellanic
Clouds]{Soszynski2015}.  Nonetheless, depending on the parent stellar
system, considerable overlap with classical pulsators of comparable
period may exist so that a clean attribution to a particular
pulsating-stars family is not always possible.  The periods of ACs
range from about 0.6 to 3~days, so that they might be confused with RR
Lyrae, BL~Her variables, or short-period Classical Cepheids.  ACs are,
however, considerably brighter than RR Lyrae stars and the lightcurves
frequently differ sufficiently that ACs are not mistaken for BL~Her
stars. ACs are either fundamental-mode (F) or first-overtone (O1)
pulsators; \emph{no} case of a double-mode pulsator is known. On the
PL plane, ACs inhabit the region between Classical and TypeII
Cepheids; with~--~at fixed period~--~being fainter than Classical but
brighter than TypeII Cepheids; also the slope of the PL relation
appears to be intermediate (cf. Fig.~\ref{fig:ACs_LMC}).  ACs are
rather rare variable stars. As mentioned before, ACs were first
identified in dwarf spheroidals, they are essentially absent in
globular clusters, and only a few are known in the Galactic field.
The study of \citet{Fiorentino2012} contains a helpful census of ACs.
The Magellanic Clouds contain the currently richest known populations
of ACs. Thanks to the monitoring efforts of the OGLE team, the
observational parameters of 141 ACs in the LMC and 109 in the SMC are
well established \citep{Soszynski2015}.  A very useful primer on ACs
with a synopsis on their history can be found in \citet{Catelan2015}.

From their locations on color~--~magnitude diagrams, mainly relative
to the horizontal branch and RR Lyrae stars thereon, and the
application of the simple period~--~mean-density relation, already
\citet{Zinn1976} estimated ACs to be two to three times more massive
than RR Lyrae stars.  The resulting conundrum of how to get stars of
the inferred mass to the observed position, say in the HR diagram, on
a time-scale in accordance with the stellar population of the hosting
stellar system, has been solved only partially.

Single-star modeling, by which observed ACs are most frequently
calibrated, calls for extremely metal-poor stars with masses of about
$0.8 - 1.8 \msol$ in the LMC \citep{Fiorentino2012}, $1.2 - 1.8 \msol$
in Scl~dSph \citep{Martinez2016}, or $0.8 - 3 \msol$ in LeoI where ACs
and short-period Cepheids are intermingled \citep{Stetson2014}.
Because ACs and Classical Cepheids are observed to follow different
period~--~luminosity relations, see e.g. \citet{Soszynski2015}, the
two families of pulsators likely obey different mass~--~luminosity
relations.  According to e.g. \citet{Fiorentino2012}, the short-period
Cepheids are interpreted as intermediate-mass stars that quietly
ignited He in their cores and which evolve through blue loops in the
HR diagram under favorable conditions (in the current context this
means mainly for sufficiently low Z-abundances). In contrast, ACs are
understood as \emph{low-mass} stars in their He core-burning phase;
such stars begin their central He burning after a series of off-center
He flashes. The locus on the HR diagram of such zero-age He
core-burning low-mass stars is found to be sensitive to their
heavy-element abundances \citep[see e.g. Fig.~3
in][]{Fiorentino2012a}: For sufficiently small $Z$~--~typically
$Z\la 0.0006$~--~zero-age He-core burning stars with masses roughly
between $1.5$ and $2 \msol$ line up along a hook that develops out of
the clump-giant region towards the classical instability strip
(CIS). The magnitude of the hook is sensitive to $Z$ and essentially
disappears for $Z > 0.001$. Stars that start their He core-burning
phase along this hook have a chance to evolve into or through the CIS
on a nuclear time-scale and hence show up as AC variables without
having to embark on a blue loop as the Classical Cepheids do.

The sensitivity on metallicity is the Achilles heel of the above
single-star AC-origin scenario.  Frequently, the postulated low $Z$
values are well below the dominating metalli\-ci\-ty of the hosting
stellar system.  This applies also to the Magellanic Clouds, both of
which host rich AC populations.  In their attempt to understand the
situation in the Large Magellanic Cloud (LMC), \citet{Fiorentino2012}
had to resort to $Z=0.0001$ and $0.0004$ models and found that such
metal-poor stars in the mass range of $0.8 - 1.8 \msol$, aged $1 - 6$
Gyrs, could feed the class of ACs therein. Hence, the invoked AC
models were considerably metal poorer than what is found to dominate
the LMC.

Binarity is not a rare configuration in stellar evolution in general
and hence potentially also in ACs. \citet{Sipahi2013a,Sipahi2013b}
reported two examples of Galactic ACs, which are binary components of
systems that must have undergone mass-transfer earlier on. Already
early on, \citet{Zinn1976} contemplated qualitatively ACs to be
possibly binary components that have passed through mass-transfer
episodes or might even be products of the coalescence of binary
components.  Detailed elaborations of scenarios do not exist, to the
best of the authors' knowledge. Nowadays, the computational
infrastructure for such an endeavor is available with MESA
\citep[cf.][]{Paxton2013,Paxton2015}.  With the robust modules for
stellar astrophysics, binary- and appropriately modified single-star
models were evolved to gain insight into the r{\^o}le binaries might
play in our understanding of ACs. The compositional choices of the
models were motivated by the best available AC population, that of the
LMC \citep{Soszynski2015}.  The results are documented in the
observer's version of the PL relation; i.e. the
period~--~Wesenheit-index (P-WI) relation: Figure~\ref{fig:ACs_LMC}
displays the 101 F-mode pulsators as filled grey circles and the 40 O1
ones as asterisks. Linear regressions to the distributions are added
as full lines for comparison with results obtained in this paper; the
respective numerical parameters are listed in Table~\ref{tab:PWI-Fits}
in the row `OGLE LMC'.  To illustrate how the ACs are positioned
relative to TypeII and Classical Cepheids, the respective F-mode P-WI
relations, again adopted from OGLE observations of the LMC, are
supplemented as dashed lines in Fig.~\ref{fig:ACs_LMC}.  Consult
\citet{Soszynski2015} for the observational background and in
particular for remarks on the attribution of pulsation modes to the
photometric data.  Throughout this paper, Figure~\ref{fig:ACs_LMC}
serves as the template against which our modeling attempts are
evaluated.
 
\section{Approach and Methods}
We investigated exemplary evolutionary paths of binary stars that are
compatible with observed ACs; the most important character traits for
us are their overstable radial pulsations in either fundamental or
first-overtone mode in the appropriate period range.  We explored
possibilities for donors, accretors, and coalesced binary remnants to
become pulsators. The performed computations and the chosen system
configurations are neither exhaustive nor possibly even
representative; they must be understood as first attempts of a
feasibility study to understand at least some of the ACs in the
context of now wide binaries that evolved into contact and lived
through mass-transfer during their earlier evolution. Binary evolution
through contact promises to resolve mismatches of age of a stellar
system and the mass of embedded stars therein.

The necessary stellar-evolution computations were all performed with
Version 8118 of the MESA software package either in single- or in
binary-star mode. In the binary mode, both components were evolved.
The binary orbits were assumed to be circular and conservative,
i.e. no mass and no angular momentum was lost from the system,
mass-loss rates were computed implicitly.  For model
stars\footnote{In MESA, the boundary of a convective core is computed by referring 
also to the chemical composition outside the boundary. 
To not systematically underestimate the duration of the He core-burning stage  
\citep[cf.][]{Bossini2015}, and to not suppress (Cepheid) loops on the HR Diagram
it is essential to allow for small overshooting at the outer edge of the convective 
helium-burning core. All MESA computations referred to in this paper had the
inlist parameter \texttt{overshoot\_f\_above\_burn\_he\_core} set to $0.001$.}
evolving into the region of the CIS, 
blue-edge locations of the lowest few orders of nonadiabatic radial pulsation 
modes were computed with Version $4.1$ of the GYRE code, which comes with the MESA package. 

Due to the lack of an implemented pulsation-convection interaction  
in GYRE, only the blue-edge loci of the instability domains were calculated in this paper. 
Any red-edges referred to in the following were ad hoc postulated by parallel shifting 
the blue edge to lower temperatures at which the periods of the pulsation 
modes were computed.
To gain confidence in the quality of the simplified pulsation
treatment, Appendix~A illustrates the computed blue edges 
of Classical Cepheid models in the LMC with neglected convective-flux perturbation. 
The theoretical F- and O1-mode 
blue edges were mapped onto the P-WI plane and compared with OGLE observations. 
The agreement is reassuring, in particular when keeping in mind that ACs are bluer
than the computed Classical Cepheids and therefore have envelopes
that are even less affected by convection.

\section{Accretor to AC}
\label{Sect:Accretors}

Deep convective stellar envelopes have the tendency to expand as a reaction to
rapid mass loss. On the other hand, if the entropy of accreted matter is not much
higher than that of the accreting stellar envelope, then it tends to
shrink upon rapid mass accretion. Therefore, simulating semi-detached binaries 
with both components on the first giant branch (FGB) undergoing case Bc mass transfer  
(i.e. case B mass transfer of donors with deep convective envelopes) 
appeared appealing: Case Bc mass transfer (MT) between two 
say $1.5 \msol$  components can be expected to strip
roughly one solar mass from the primary and feeding it onto the secondary, transforming 
the latter to an intermediate-mass star, which takes a much longer time to evolve into Helium 
core-burning than an isolated star starting from the ZAMS with the same mass. 
The property of the accretor to shrink prevents it from also filling its Roche lobe 
and from forcing the system into contact.
Even though the remnants stay bound, their orbit will widen so much that the binary 
nature of a pulsating component might be difficult to establish observationally.

The strong mass dependence of stellar main-sequence lifetime necessitates small
mass differences between the components of binaries aimed at. The set of computed systems 
consisted of almost-twins on the ZAMS; the primary was usually about $0.02 \msol$  
more massive than the secondary. A larger mass difference left the secondary
too far back in its evolution, usually with the secondary still having a predominantly 
radiative envelope at the onset of case Bc MT and therefore leading to contact systems, 
which cannot be dealt with in MESA.
The smaller the mass differences the more pronounced is the tendency of both components
to fill their respective Roche radii at the same time  so that transferring enough mass 
before full contact becomes difficult again. 

For our exploratory purposes we evolved a few almost twin-component 
binaries with $Y=0.3, Z=0.001$ through case Bc MT until the accretor evolved to the
end of He core-burning. We restricted the ZAMS masses of the component so that the accretor mass 
was $< 3 \msol$ at the end of MT. The initial separation of the binary components on the ZAMS 
varied between $80 - 120 \rsol$ (equivalent to about $50 - 100$~d orbital period), 
this ensured MT to start during the FGB phase of the donor 
and the accretor having already started its ascent along the FGB 
(in case the mass difference between the components was $\la 0.02 \msol$).
We found the post-MT masses to not depend crucially on the size of the orbit, therefore
we performed most binary computations with an initial separation of $100 \rsol$.

\begin{figure}
	\includegraphics[width=\columnwidth]{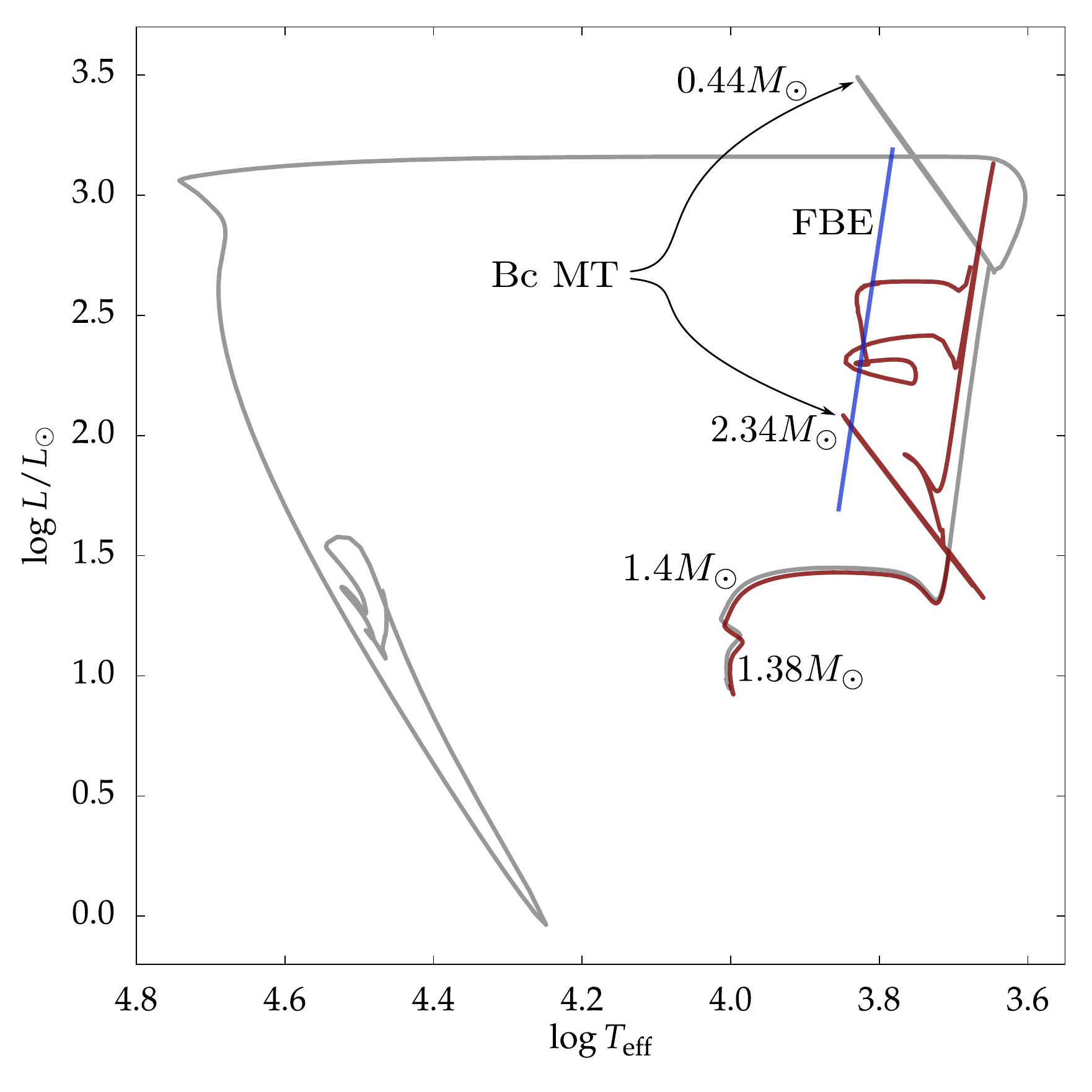}
    \caption{Illustrative evolutionary tracks of a binary system with $1.38$ and 
    	     $1.40 \msol$ components with initially $100\rsol$ separation and 
    	     the abundances $Y=0.3, Z=0.001$. Unstable mass transfer of case 
    	     Bc (Bc~MT) is encountered when both components evolve along the FGB.
    	     The locus of the fundamental-mode blue edge (FBE) is sketched 
    	     as a blue line running roughly parallel to the FGB loci 
    	     of the binary components.
    	    }
    \label{fig:Bc_Evol}
\end{figure}

\begin{figure}
	\includegraphics[width=\columnwidth]{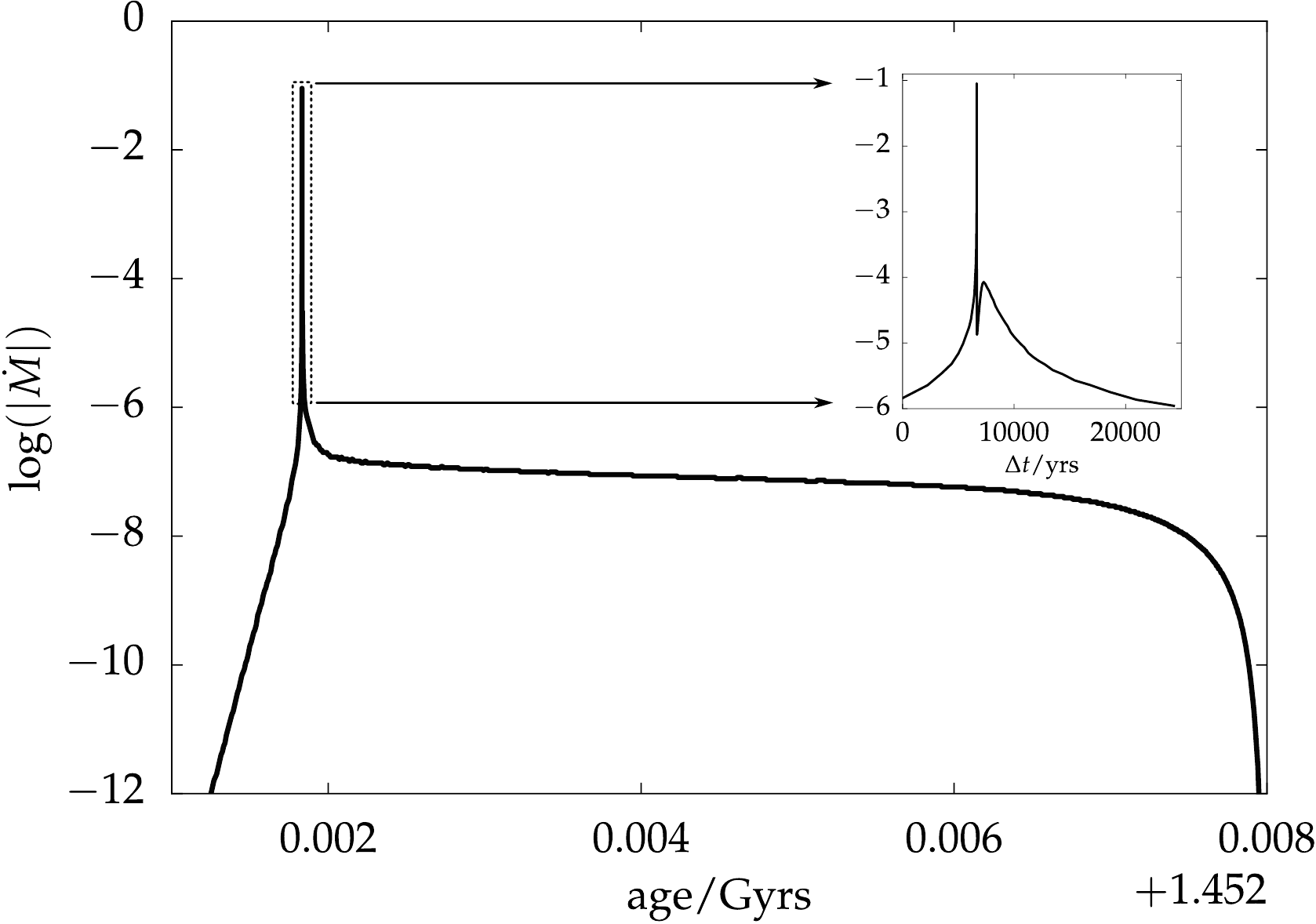}
    \caption{Mass-transfer history for the $1.4/1.38\,\msol,\,100 \rsol$-separation binary model 
    		 shown in Figure~\ref{fig:Bc_Evol}. The inset to the right zooms in to the
    		 phase when $\dot{M} > 10^{-6} \msol\mathrm{yr}^{-1}$. 
    		 The very narrow initial peak, when
    		 the mass-transfer rate approaches $10^{-1} \msol \mathrm{yr}^{-1}$, 
    		 spans only a few years of the stars' evolution.
    	    }
    \label{fig:Bc_Evol_MT}
\end{figure}

Figure~\ref{fig:Bc_Evol} illustrates representative evolutionary tracks of binary components
with initially $1.4$ (dark/red locus) and $1.38 \msol$ (brighter grey locus), respectively, with an orbital period of $69$~days, corresponding to a separation of $100 \rsol$, on the ZAMS.
The tracks of both binary components look essentially identical to the lower FGB 
when dynamically unstable MT sets in; the respective reactions of the two components 
on the HR plane are highlighted as `Bc MT'. The short episode of case Bc MT shows up as 
diagonal excursions traced out by the donor as well as the accretor.
Figure~\ref{fig:Bc_Evol_MT} shows the details of the evolution of the 
mass transfer rate during the Bc MT episode, which lasts for about 7 million years. The phase
of $\dot{M} > 10^{-6} \msol \mathrm{yr}^{-1}$ persists for roughly $24\,300$~yrs, the one during 
which most of the mass is transferred from the primary to the secondary, 
when $\dot{M} > 10^{-3} \msol \mathrm{yr}^{-1}$, spans about 
$6.5$ years. Both components involved in the Bc MT process show strong luminosity and
effective temperature variations, essentially at constant radius, on the HR plane
(see Fig.~\ref{fig:Bc_Evol}). It remains unclear if this behavior is an artifact of the 
numerical implementation of the dynamically unstable mass transfer. 
The ensuing evolution of the remnants of Bc MT remained robust, independent of 
the particular settings of the numerical treatment of this evolutionary phase.

During case Bc MT, the primary lost $0.96 \msol$ to the secondary. 
The resulting $0.44 \msol$ remnant still ignited helium burning and evolved 
into a low-mass white dwarf, which eventually underwent a  late shell flash 
(cf.~Fig.~\ref{fig:Bc_Evol}).
The secondary grew from $1.38 \msol$ to $2.34 \msol$ within a few tens of thousands of years, 
i.e. on a time-scale without noticeable nuclear evolution. The thermal relaxation of the new star 
can be seen in the HR diagram of Fig.~\ref{fig:Bc_Evol} as the hook that developed above the 
Bc MT fault of the evolutionary track (at around $\log \llsol = 1.75$). Afterwards, the by then 
more massive secondary continued to evolve along the FGB in quasi-equilibrium until it started 
its helium core-burning. Even though the newly minted primary qualified as an 
intermediate-mass star which normally starts helium core-burning quietly, it continued 
to behave like a low-mass star in its center. Helium core-burning started 
degenerately in a series of mild off-center flashes during which the star evolved off the tip 
of the  FGB  to the region of the FBE at about $\log\llsol = 2.25$. 
Hence, we could indeed produce radially pulsating He-burning stars with periods short enough 
to be AC candidates as the result of some close-binary MT episode. 
By the time the more massive secondary passed through the CIS it was accompanied
by a low-mass binary companion that was about 30 times fainter and orbited the pulsating star 
with a period of around 420 days.

\begin{figure}
	\includegraphics[width=\columnwidth]{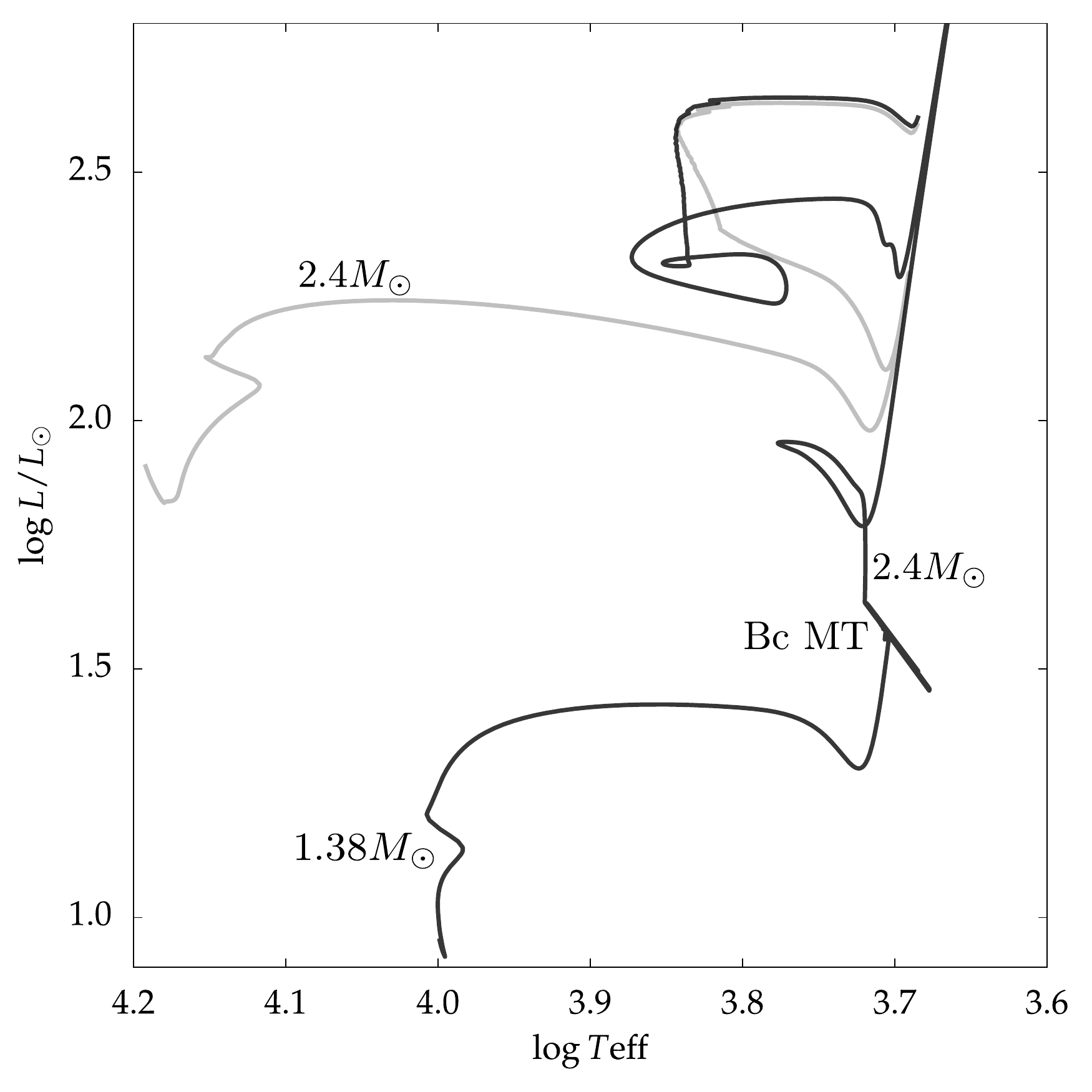}
    \caption{Approximating the accretor component via a single-star 
    		 evolution approach (details in the text) is plotted with the black line.
    		 The grey locus shows the evolution of a $2.4 \msol$
    		 model star from the ZAMS in absence of any mass accretion. }
    \label{fig:Bc_Single}
\end{figure}

\subsection{Emulating case Bc with single-star evolution}
\label{sect:SingleStar_Bc}
\begin{figure}
	\includegraphics[width=\columnwidth]{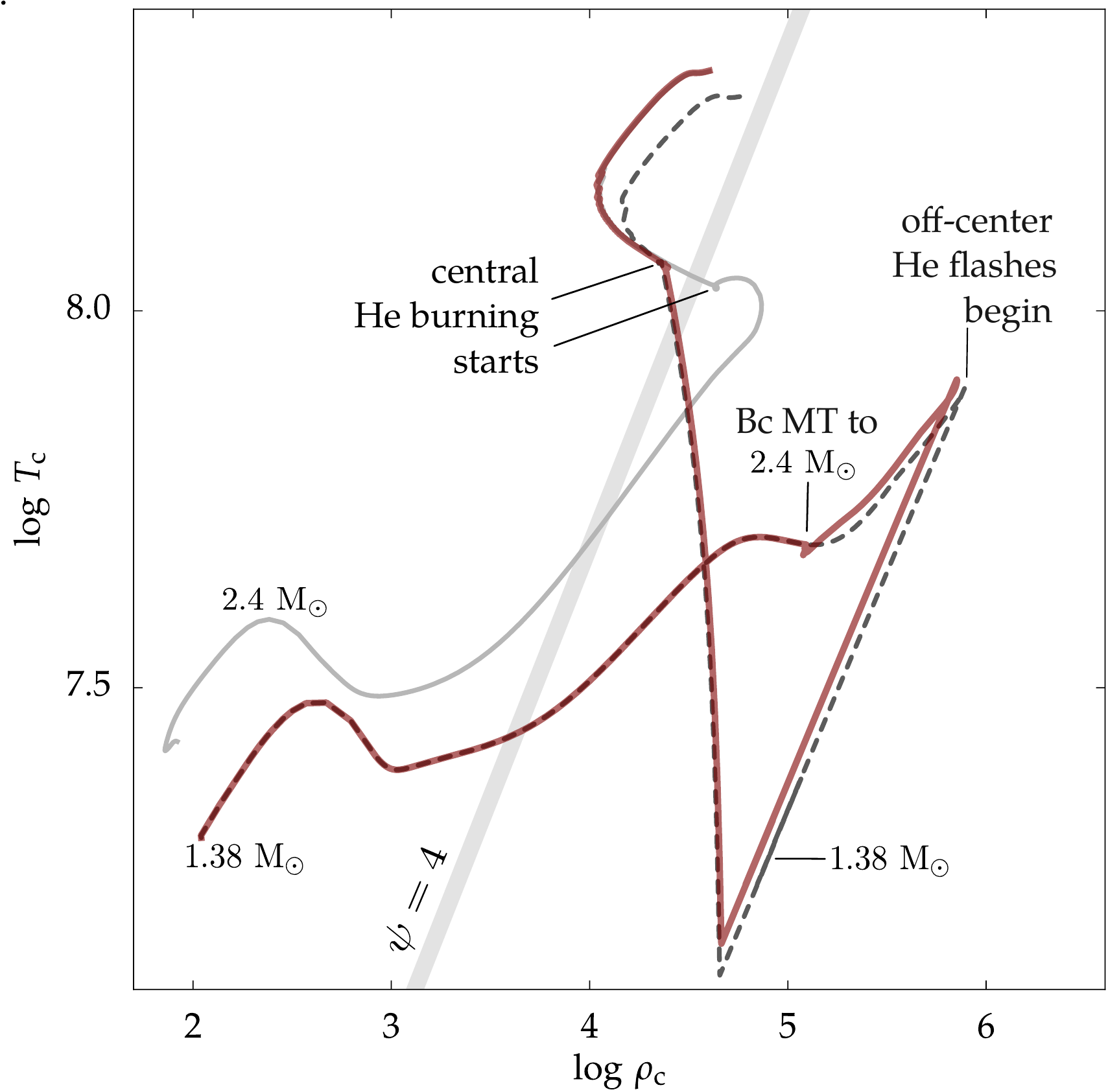}
    \caption{The evolution of central properties of model stars with $1.38$ and $2.4 \msol$
    	     computed without tampering with their mass is plotted in grey dashed 
    	     and full lines, respectively. The behavior of the central conditions of the 
    	     initially $1.38 \msol$ star, which gains 
    	     $1.02 \msol$ via Bc MT like mass accretion is traced out by the red line.  
    	    }
    \label{fig:Bc_SingleRhoT}
\end{figure}

The very short mass-transfer episode and, except for mass accretion, the negligence 
of any other binary-specific physical interaction invites an emulation of accretor
evolution by means of a single-star ansatz. Figure~\ref{fig:Bc_Single} illustrates the
outcome of the evolution of an initially $1.38 \msol; Y=0.3, Z=0.001$ star from the ZAMS. 
It underwent rapid mass accretion on the lower FGB until it reached, in accordance with 
the binary system presented above, a total mass of $2.4 \msol$. The resulting evolutionary track
is shown by the black line. The epoch of rapid mass transfer, which was set constant to 
$\dot{M} = 10^{-3} \msol \mathrm{yr}^{-1}$, is indicated again with `Bc MT'. 
The agreement of the $1.38/2.4 \msol$ loci in Fig.~\ref{fig:Bc_Evol} and Fig.~\ref{fig:Bc_Single} 
is comforting. To emphasize the differences of intermediate-mass stars that 
evolved as isolated stars from the ZAMS  and those which turned into 
intermediate-mass stars later in life via close-binary interaction, 
a $2.4 \msol$ star with $Y=0.3, Z=0.001$ was evolved from the ZAMS to the end of He core-burning. 
The respective evolutionary track is  shown by the grey line in Fig.~\ref{fig:Bc_Single}. 
Despite their divergent early phases, the tracks 
of the two $2.4 \msol$ model stars converged on the HR diagram  
during He core-burning and they ended up essentially congruent at the
beginning of double-shell burning. A corresponding behavior was also observed on 
the $\log \rho - \log T$ plane of the stars' centers in Fig.~\ref{fig:Bc_SingleRhoT}: 
The $2.4 \msol$ track shown in light grey, the track described by a 
comparison star of $1.38 \msol$ without mass transfer is plotted with a dashed line; 
finally, the evolution of central density and temperature of the initially $1.38 \msol$ star
that gains $1.02 \msol$ mass in a Bc-type MT along the FGB is colored in red 
(which appears as a darker grey in the printed version). 
Evolution starting on the ZAMS places the three models on the lower left of the diagram. 
The broad grey band crossing diagonally
through the plot constitutes a line of constant degeneracy, 
chosen to be at $\psi=4$. The weak degeneracy of the stellar matter in the center of 
the $2.4 \msol$ model does not destroy the thermal stability of central He-burning 
at ignition: i.e. the  $2.4 \msol$ star starts central $3\alpha$ burning quietly.
The $1.38 \msol$ star's central evolution is also textbook-like: After central H-burning, 
central temperature and density rise continuously, with the density getting high enough
for stellar matter going degenerate. At the tip of the FGB, the $1.38 \msol$ model star
develops off-center He-flashes. In a series of thermal instabilities $3\alpha$ burning propagates
to the center where thermal stability is eventually regained and quiet central He-burning takes
over. At that epoch the evolutionary loci on the  $\log \rho_{\mathrm{c}} - \log T_{\mathrm{c}}$ 
plane of the $1.38 \msol$ and the $2.4 \msol$  models convene. 
A discussion of the wedge-shaped locus associated with the off-center He-flashes can
be found e.g. in \citet{Paxton2011} or \citet{Gautschy2012}.
The ensuing core He-burning phase for both models are similar, except that density was slightly
higher but the temperature lower at the center of the $1.38 \msol$ star.
Focus now on the accretor track: It starts as a $1.38 \msol$ star and gains quickly
$1.02 \msol$ of matter; the corresponding epoch is marked in 
Fig.~\ref{fig:Bc_SingleRhoT} as `Bc MT to $2.4 \msol$'. 
On the HR diagram, the thermal relaxation, after the Bc MT episode,  
of the then more massive star showed up as a small blue loop only at around $\log \llsol = 1.8$ in 
Fig.~\ref{fig:Bc_Single}. Even though the accreted mass is considerable, the effect on the
further evolution of the star on the $\log \rho_{\mathrm{c}} - \log T_{\mathrm{c}}$ plane was 
small. It is the sharp density contrast across the core~--~envelope boundary that essentially 
decouples the two regions so that the core's evolution does not depend 
sensitively on what happens to the tenuous outer layers \citep[cf.][]{Sugimoto1980}. 
The newly minted $2.4 \msol$ star continued to behave very much like the unperturbed
$1.38 \msol$ star. In particular, also the $2.4 \msol$  accretor developed a thermally unstable 
He-shell and core He-burning is established via a series of He-shell flashes. 
The final part of the wedge-shape locus leading out of degeneracy is identical for 
the accretor and for the $1.38 \msol$ star. However, once core He-burning took over the 
accretor switched over to follow the locus of the unperturbed $2.4 \msol$ star.

On the HR diagram, it is evident that the degenerate onset of He core-burning 
takes the accretor to the blue earlier than 
in the case of quiet nondegenerate ignition. Therefore, the star evolved into and through 
the CIS at lower luminosities; i.e.~after degenerate ignition of core-burning
accretor remnants  have a mass~--~luminosity relation that differs from canonical 
nondegenerate igniters.

\begin{figure}
	\includegraphics[width=\columnwidth]{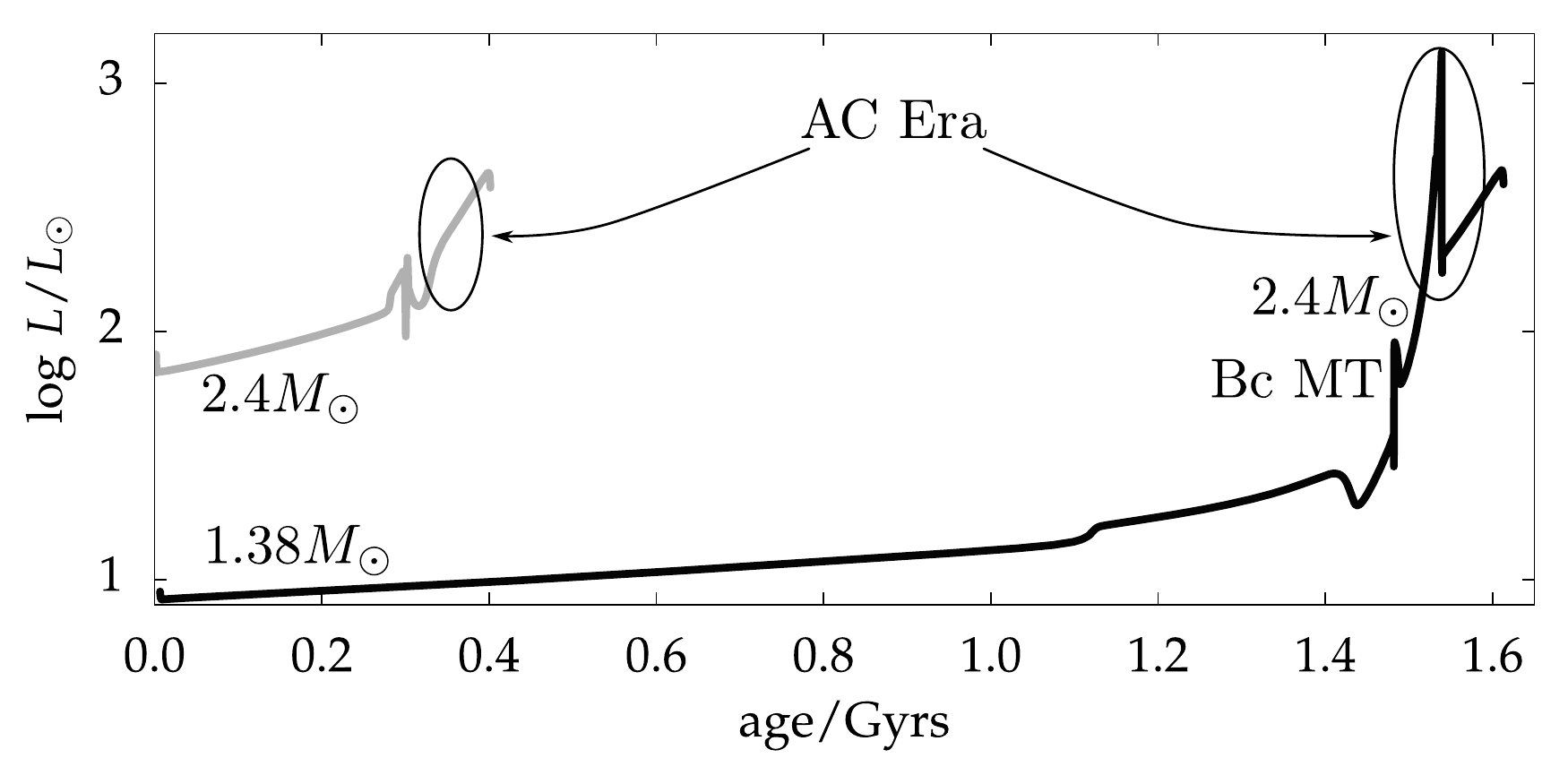}
    \caption{Evolutionary retardation as effected through close binarity, illustrated with the 
    	 	 two model sequences shown in Fig.~\ref{fig:Bc_Single}. 
    	    }
    \label{fig:Bc_Retardation}
\end{figure}

Another important effect of Bc MT on the mass-gainer is the evolutionary retardation 
effect inflicted by the process. Figure~\ref{fig:Bc_Retardation} plots age in gigayears 
versus luminosity. The pertinent epochs for the AC phenomenon are highlighted by ellipses. 
An isolated $2.4 \msol$ star reached its AC domain after about
0.35 gigayears. In contrast, the initial $1.38 \msol$ star that quickly gained $0.96 \msol$ along 
the FGB took about 1.5 gigayears to become a potential AC. 
In other words, binarity helps to accommodate seemingly too massive stars in otherwise
older stellar populations. 

\subsection{Pulsation properties of case Bc MT accretors}
\begin{figure}
	\includegraphics[width=\columnwidth]{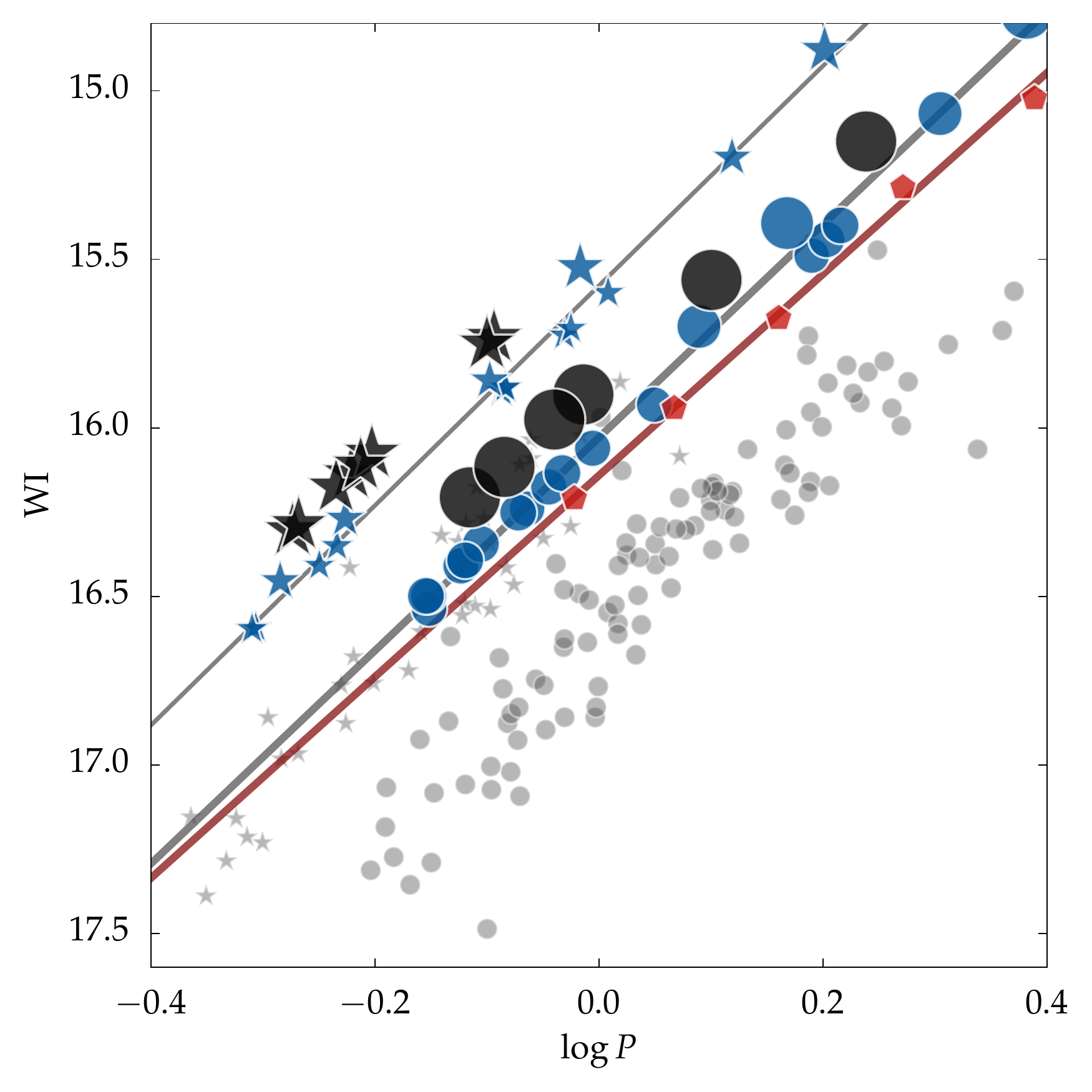}
    \caption{P-WI correlation obtained for the 
    		 accretor models of case Bc mass-transfer binaries 
    		 with $Y=0.3, Z=0.001$ (blue). Black (big) markers show the
    		 results for the $Y=0.3, Z=0.003$ sequence.
    		 The circles indicate fundamental-mode
    		 pulsators, the asterisks first-overtone pulsators at their
    		 respective pulsation blue edges. The size of the markers
    		 scales with the mass of the pulsating star. Row `Case Bc-A' of 
    		 Table~\ref{tab:PWI-Fits} contains
             the numerical values of the linear-fit parameters.
             Pentagons and the fitted red line illustrate the effect of
             the finite-width of the CIS on the P-WI diagram; details are given in the text.}
    \label{fig:PWI_BcA}
\end{figure}

For the ad hoc chosen chemical composition of $Y=0.3$ and $Z=0.001$ we computed a number of 
evolutionary tracks either tracing the accretor in MESA's binary mode, or then~--~when we
studied model stars with different chemical compositions~--~to speed up the process,
following the procedure described in Section~\ref{sect:SingleStar_Bc}. 
Figure~\ref{fig:PWI_BcA} displays the results at the encountered blue edges with blue 
(dark grey in print) markers: Fundamental modes are shown as filled circles and first overtone
modes as asterisks. For comparison, the computational results are overplotted on the 
LMC observations in light grey.
The models' luminosities were transformed to absolute photometric $V$ and $I$ magnitudes using a 
public FORTRAN program with an implementation of fits of stellar parameters to photometric
passbands by \citet{Worthey2011}. To eventually dislocate the models to the right distance, 
an LMC distance modulus of $18.5$ mag was applied.  

The grey lines through circles and asterisks plotted in Fig.~\ref{fig:PWI_BcA} are linear 
least-square fits to the computed blue-edge locations, therefore they should prescribe the upper
envelope to the respective observational data.  
To get an idea of the effect of the finite width of the CIS on the P-WI diagram, 
we computed F-modes at positions along the respective evolutionary tracks 
red-shifted by $0.05$ in $\log \teff$ from the FBE positions.
Evidently, the instability strip collapses to a high degree on the P-WI plane; 
i.e. modeling predicts the intrinsic width of the instability strip on the P-WI plane 
to be considerably narrower than what observed fundamental mode ACs in the LMC suggest.
The effect of chemical composition on the blue edges is illustrated with the additional 
black markers (which are also bigger) in  Fig.~\ref{fig:PWI_BcA}; the respective model sequence 
used  $Y=0.3$ and $Z=0.003$ and was computed in single-star approach with imposed 
fast MT on the FGB. 
A slight metallicity dependence of the P-WI relation seems to obtain: 
At a given period, the blue edges of the more metal-rich models are slightly brighter.

The size of the CIS~--~blue-edge markers in the P-WI diagrams 
scales with the mass of the model stars. 
The $Y=0.3$ and $Z=0.001$ models entering the CIS had accretor remnants with 
$2.3 < M/\msol < 2.8$  for the chosen binary parameters 
(such as initial mass of the components and initial separation). 
The general trend, which is though not strict, is that lower-mass stars populate
the shorter-period and fainter WI domain of the plane, this applies to fundamental 
and first-overtone modes. 
In case of  $Y=0.3$ and $Z=0.003$ models, $3 \msol$ is the lowest final accretor mass 
which crossed the CIS. Apparently, depending on its looping properties, even a $3 \msol$ star 
can pop up almost anywhere in the period range occupied by observed ACs.

\section{Donor to AC}
In last section we succeeded to evolve accreting components of interacting binaries 
into ACs. Therefore, the same can be conjectured to be possible for donors. 
To speed things up, we computed single-star models from which, in a 
case Bc-like scenario, mass was abstracted at the rate of  $10^{-3} \msol \mathrm{yr}^{-1}$
once a prescribed helium core-mass was reached, 
i.e. once the model star expanded to an appropriate radius on the FGB. The adopted mass-loss rate 
is arbitrarily set, with robust results, as long as it is large enough for the remnant 
mass (also prescribed) to be reached on a
timescale shorter than the star's Kelvin-Helmholtz time; this ansatz emulates a dynamically 
unstable mass transfer. The price to pay is that we cannot follow the behavior of the binary system 
such as the separation of the components and the actual stop condition of the case Bc MT. 
In other words, more careful studies are required to find actual binary \emph{system} configurations 
that reproduce donor stripping as prescribed in the ad-hoc way of this section.

\begin{figure}
	\includegraphics[width=\columnwidth]{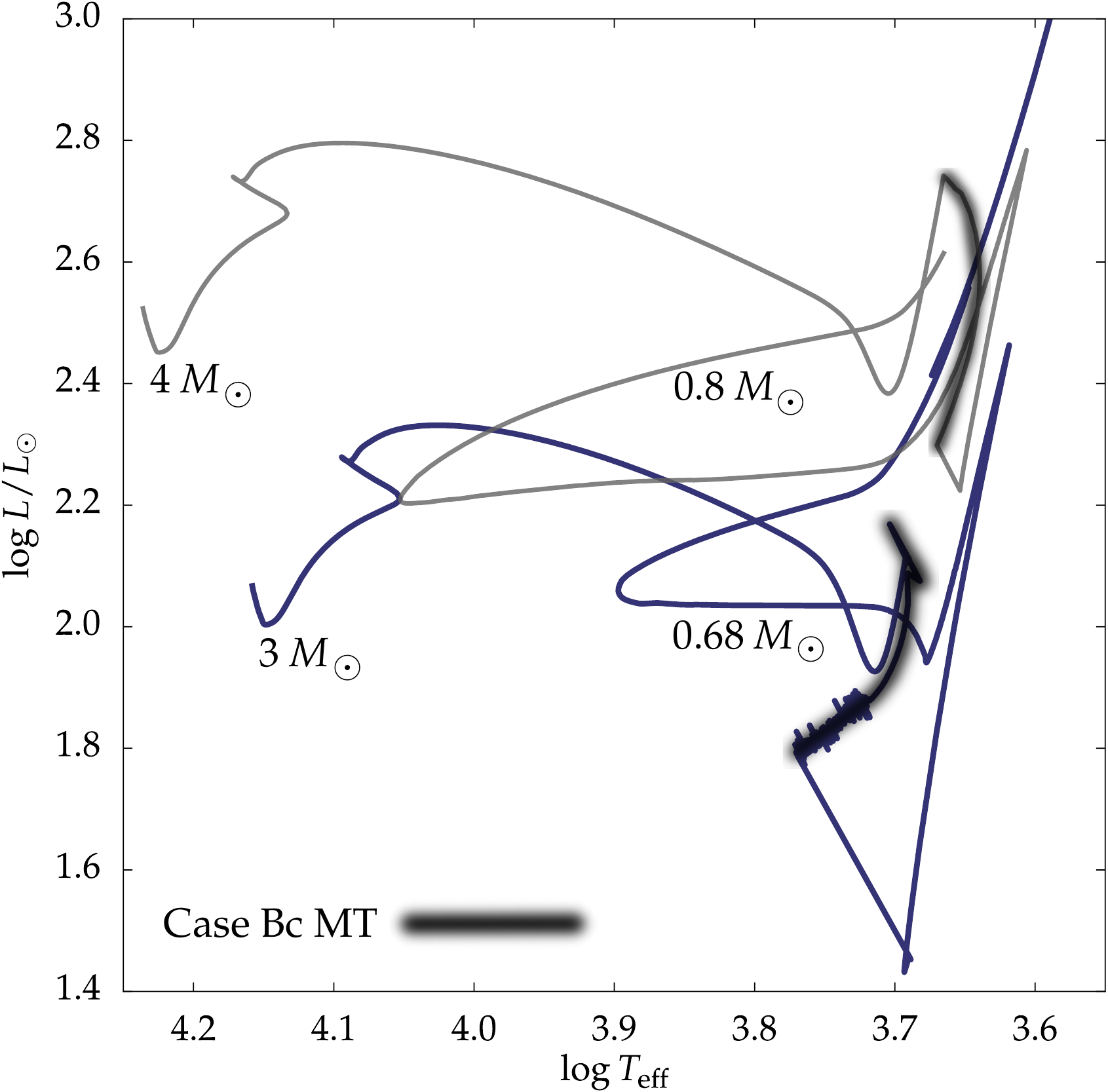}
    \caption{Evolutionary tracks of representative, initially $3$ and $4 \msol$
    		 stars from ZAMS to the end of He core-burning. Case Bc MT along the FBG, 
    		 reduced the stellar masses to $0.68$ and $0.8 \msol$, respectively. }
    \label{fig:Bc_Donor}
\end{figure}

Stars of initially $3.0, 3.5, 4.0, 4.5$, and $5.0 \msol$ and $Y=0.254, Z=0.006$ 
were evolved off the ZAMS. For suitable prescriptions of core He- and total remnant masses, 
fast MT along the FGB led indeed to CIS-crossing stars with remnant masses 
in the range $0.6 - 1.06 \msol$. Two examples are shown in Fig.~\ref{fig:Bc_Donor}.
Note that the donor models began their evolution as intermediate-mass stars which then 
turned into low-mass stars along the FGB. Much in accordance with the 
findings in the accretor case (cf. Sect.~\ref{Sect:Accretors}), also the donor remnants continued 
their evolution on $\log \rho_{\mathrm{c}} - \log T_{\mathrm{c}}$ plane along the path of the model
star \emph{before} case Bc MT; in this case, evolution of the central quantities of the 
model stars continued just like that of intermediate-mass stars even though the actual models 
had lost most of their initial mass. For the low-mass donor remnants this meant that they 
started their helium core-burning quietly under only weakly degenerate conditions. 
Depending on the choices of He core-mass at Bc MT switch-on and on the targeted 
remnant mass, we found that considerably He-enhanced envelopes could develop. 
In other words, ACs resulting from stripped donors might give away their 
origin spectroscopically via higher He/H ratios.

\begin{figure}
	\includegraphics[width=\columnwidth]{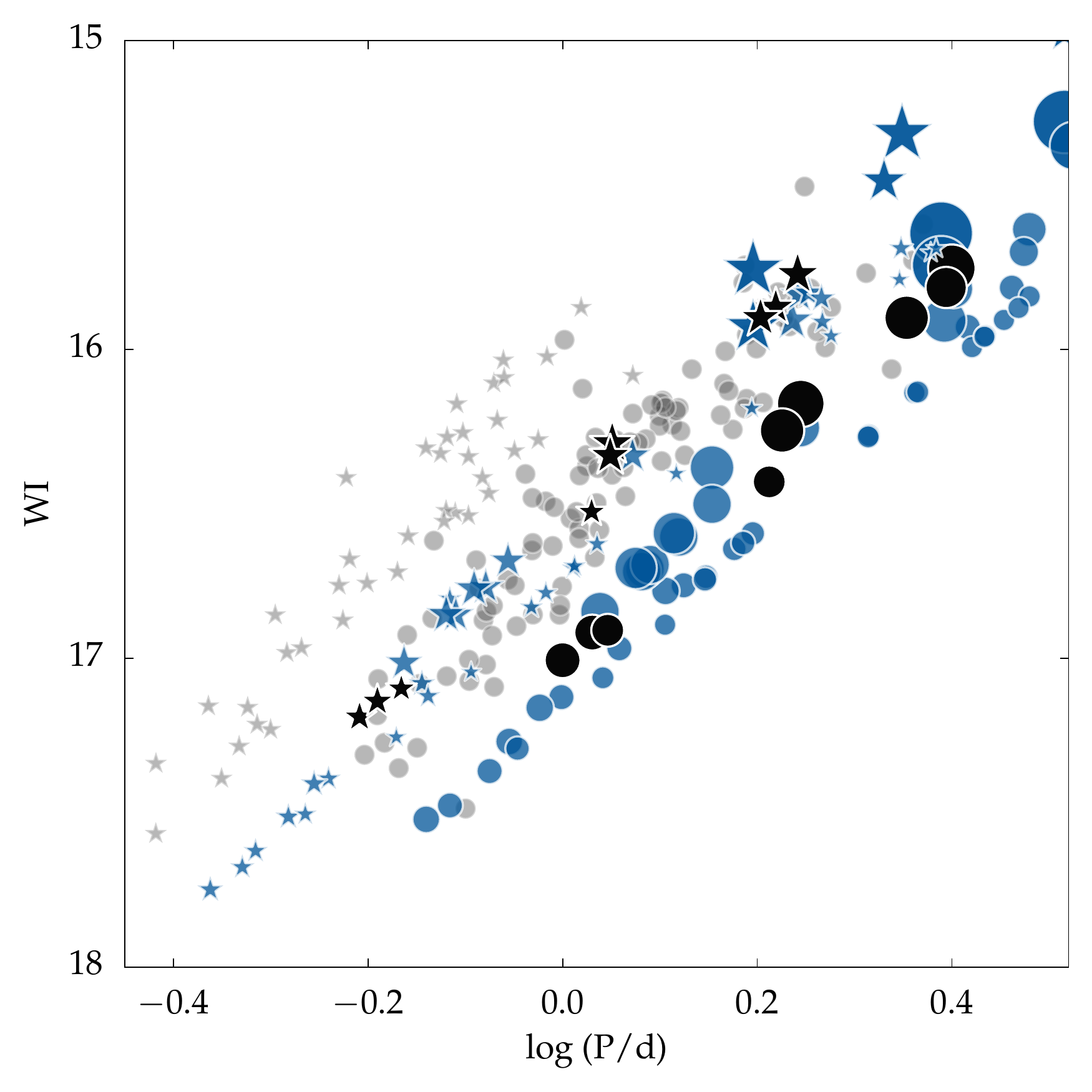}
    \caption{The P-WI correlation emerging from  
    		 donor models in case Bc mass-transfer binaries 
    		 with $Y=0.254, Z=0.006$ (blue) and $Z=0.008$ (black, $Y$ unchanged). 
    		 }
    \label{fig:PWI_BcD}
\end{figure}

In Fig.~\ref{fig:Bc_Donor}, two stellar evolution tracks illustrate representatively 
the behavior of donor models. An initially $3 \msol$ star was stripped of its 
envelope on the FGB when its He core-mass reached $0.35 \msol$; 
a very high case-Bc~--~like mass loss removed mass until the remnant 
was left with $0.68 \msol$. As evident from the dark blue evolutionary track, 
the remnant set out on a blue loop, crossing the CIS, during He core-burning. 
The same general behavior, with an even more extensive blue loop, 
applies to the $4 \msol$ model star, whose mass loss started at 
a He core mass of $0.49 \msol$. Mass was abstracted until the remnant was left with $0.8 \msol$. 
The phase of rapid mass loss is emphasized with a thick foggy line. 
The straight loci to lower luminosities and effective temperatures after termination of rapid mass 
loss were traced out when the remnants approached thermal equilibrium again. 

The results on the P-WI plane of the blue edges 
as computed from our donor-stripping experiments are displayed in Fig.~\ref{fig:PWI_BcD}.
Most importantly, from stars with initial masses exceeding $3 \msol$ and for heavy 
element abundances as high as $Z=0.006$ and $Z=0.008$ remnants were found to enter 
and cross the CIS with pulsation periods that are compatible with ACs. The weak grey 
background data points in Fig.~\ref{fig:PWI_BcD} are again those 
observed of ACs in the LMC.

The mass of the models at their F and O1 blue edges are encoded again in the size of the markers. 
The smallest ones represent $0.6 \msol$, the biggest ones $1.06 \msol$ pulsators. The markers
in blue (intermediate grey) indicate the $Y=0.254, Z=0.006$ sequences. 
For initial masses of $4.0$ and $4.5 \msol$ 
and a heavy-element abundance of $Z=0.008$ (at fixed $Y=0.254$) blue looping 
helium core-burning models were found for remnant masses between $0.73$ and $0.88 \msol$; 
their positions on the P-WI plane are shown by the black symbols in Fig.~\ref{fig:PWI_BcD}. 
Within the scattered blue-edge data, no obvious metallicity dependence can be perceived.
Furthermore, more massive pulsators are found at longer periods and higher luminosities. The
lower-mass kin, however, are not restricted to short period and lower luminosity; they 
scatter along the whole AC-relevant period range of $0.6 - 2.4$~days of F modes 
(and $0.4 - 1.6$~days of O1 modes). It can even happen that small-mass O1 modes occupy 
the same region  on the P-WI plane as more massive F-mode ACs. 
One can only hope that such a situation  can be disentangled with the help of differing 
lightcurve shapes.    

Because the scatter of even the blue-edge data is considerable we refrained from a fit 
to quantify the P-WI relations. Nonetheless, the situation in 
Fig.~\ref{fig:PWI_BcD} is clear: The computed blue-edge data 
(adopting again a distance modulus of $18.5$~mag) for F and O1 modes 
are roughly parallel shifted to the observed LMC P-WI data. In contrast to Fig.~\ref{fig:PWI_BcA},
the ACs from donor remnants in Fig.~\ref{fig:PWI_BcD} appear to be about 0.5~mag fainter than what 
LMC observations require.

The donors-to-ACs scheme can produce the sought-of pulsators in about 
$4\times 10^{8}$~yrs (for $3 \msol$ stars on the ZAMS) and $1\times10^{8}$~yrs 
(for $4.5 \msol$ ZAMS stars). Donor remnants turning into ACs would hence require 
star formation within the last few hundred million years in the hosting stellar system. 

\begin{figure}
	\includegraphics[width=\columnwidth]{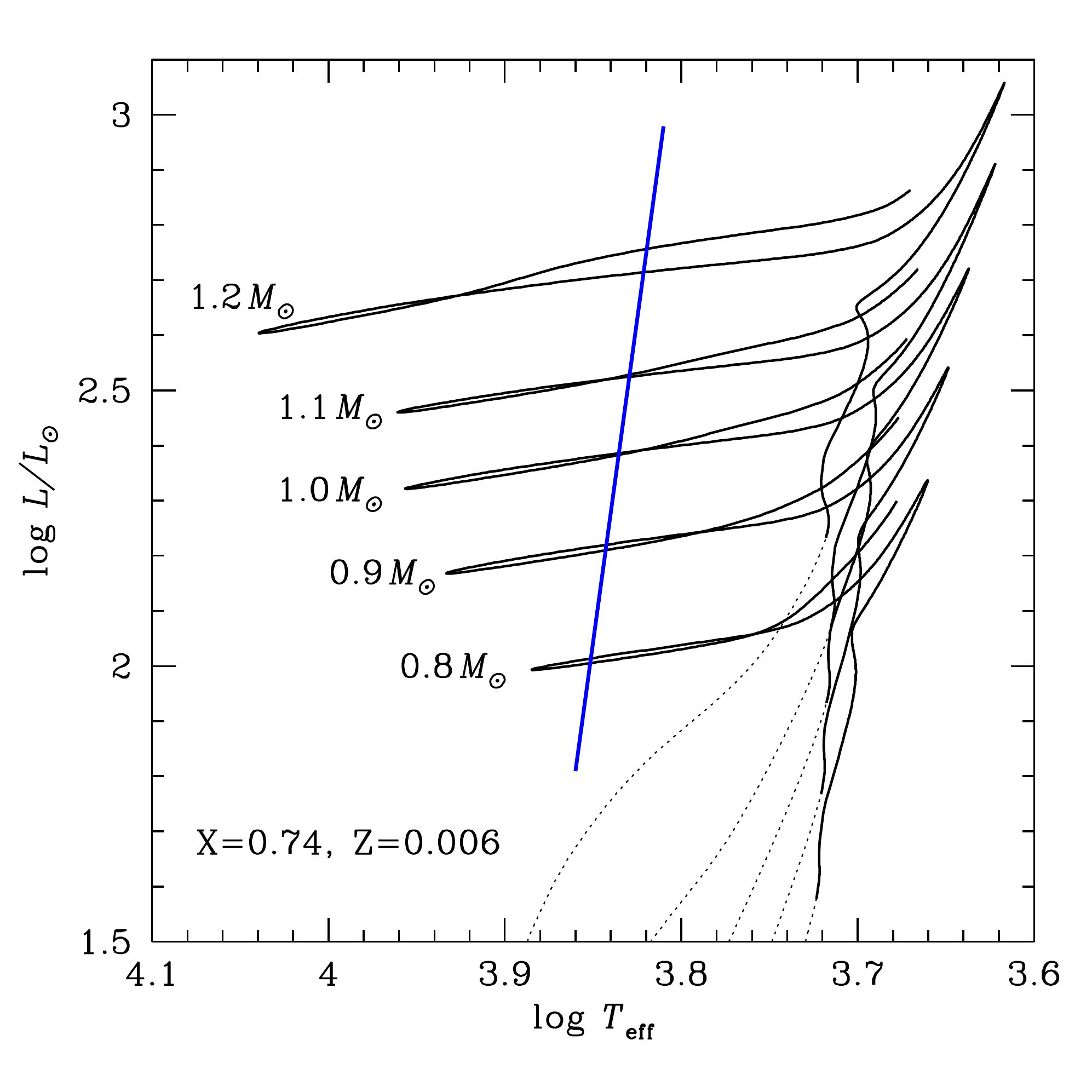}
    \caption{Evolutionary tracks of merger models with $Y=0.254$ and $Z=0.006$ that extend 
    		 to exhaustion of He in the nuclear burning center. 
    		 Hydrogen core-burning prevails along the dotted part of the loci; 
             for the merger models discussed here, this stage is not relevant.
     		 We assume that the object emerging from the coalescence of two
     		 binary components starts its single-star evolution with a non-degenerate core  
     		 somewhere before the onset of He core-burning.
     		 Helium in the stars' centers ignite around maximum luminosity of the displayed 
     		 evolutionary tracks. The straight blue line at the center of the figure 
     		 traces the location of the FBE.    
    	    }
    \label{fig:hrd_merge}
\end{figure}

\begin{figure}
	\includegraphics[width=\columnwidth]{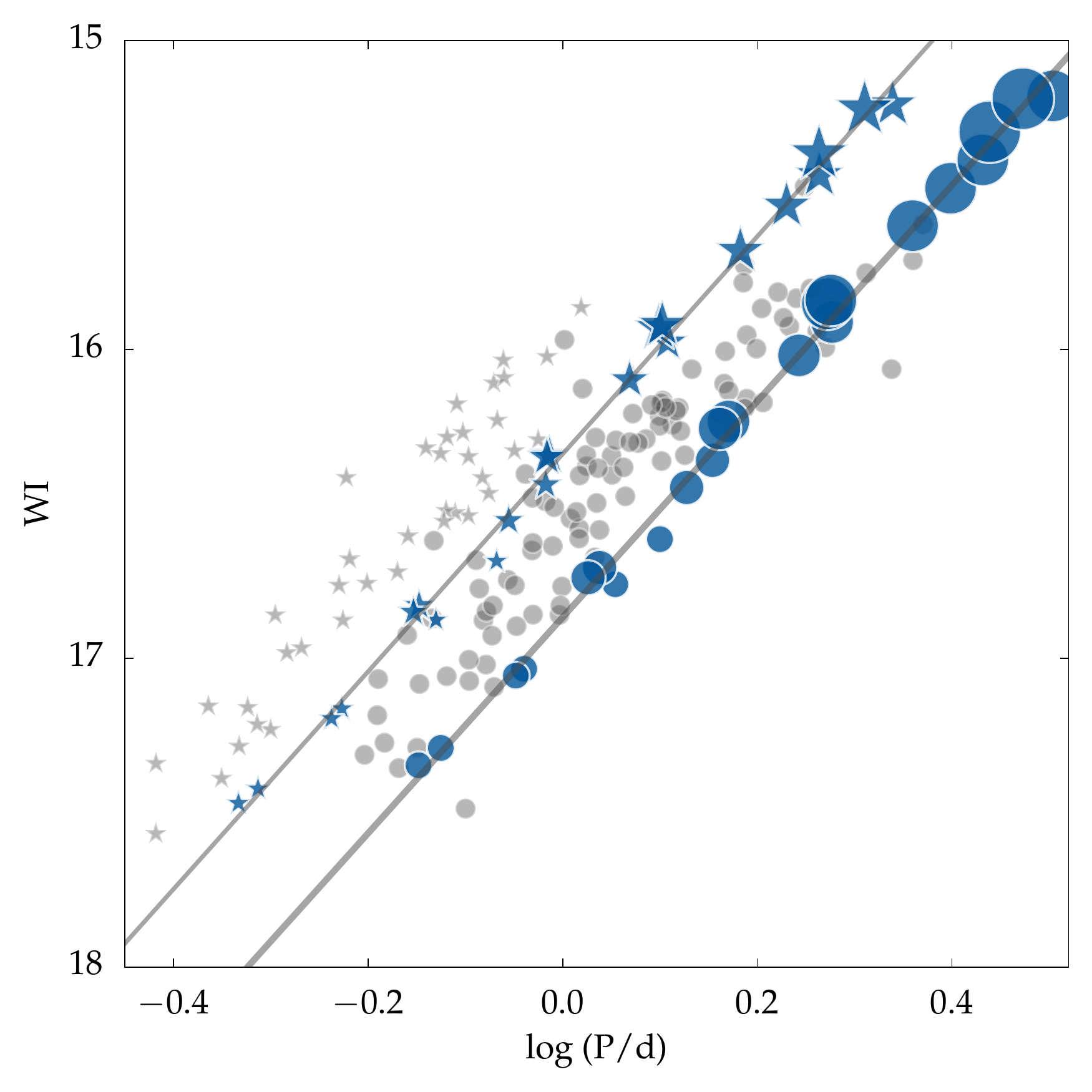}
    \caption{The P-WI diagram as obtained from merger emulations. The blue-edge locations
    	    are plotted for fundamental and first-overtone modes  
    	     with $Y=0.254$ and $Z=0.006$. 
    	     The positions of the blue edges were found to change on $Z$ modifications 
    	     only marginally.   
    	    }
    \label{fig:PWI_Mrgr}
\end{figure}

\section{Merger to AC}
Finally, we conjectured that with some binary parameters mass transfer leads 
to a merger of the  involved stellar components. 
In the following, we consider 
possibilities for such coalesced configurations to evolve through the CIS as ACs.
The generic property of a post~--~main-sequence star on the FGB is its well 
expressed `core-halo structure'; i.e. a dense radiative He core smaller than about $0.1\,\rsol$  
with a tenuous hydrogen-rich envelope extending as far as a few times $10\,\rsol$. 
Because of the large density contrast between core and envelope, the core structure of such a 
star is insensitive to the envelope mass.
Therefore, we expect the structures of the two cores to be hardly affected 
even upon the formation of a common envelope, and also once the cores coalesce 
in such a common-envelope scenario. The details of such a phenomenon is a complicated dynamical,
three-dimensional process. All we can do
here is to speculate that the emergent object is a star with a helium core whose mass is about 
a sum of the two initial cores. The remaining H-rich envelope mass, on the other hand, is 
supposedly significantly smaller than the sum of the masses of the original envelopes because 
a large amount of the tenuous envelopes is likely to be expelled from the system 
during the merging process \citep[e.g.][and references therein]{Ivanova2013}.

To obtain rough one-dimensional models that represent what we expect to emerge 
at the end of a coalescence and to emulate their evolution through the ensuing 
He core-burning stage, we computed, resorting again to MESA, respective 
model stars with $0.8 - 1.2\,\msol$. To approximate the expected partial 
mixing around the merging cores on their way to become about twice as massive 
as that of normal evolution, extensive overshooting from the H-burning core was invoked.  
Overshooting was computed with the prescription of exponentially decaying mixing 
as implemented in MESA, in which a parameter $h_{\rm ov}$  measures the 
scale-length ($h_{\rm ov}H_p$ with $H_p$ being the pressure scale-height) of the exponential 
decay of the mixing efficiency above the convective core \citep{Herwig2000}. 
The quantity  $h_{\rm ov}$ was varied in the range of $0.22 - 0.25$ with the aim to 
obtain evolutionary tracks that take the stars through the CIS.  Appropriate values 
of $h_{\rm ov}$ were found by trial-and-error, whereas larger values were required 
for smaller-mass models. With the very large overshooting as adopted here, He core masses between 
$0.6$ and $0.9 \,\msol$ resulted so that helium ignited non-degenerately at the center.  
Extended loops occurred during He core-burning if the ratio of envelope mass to the total 
mass was about $0.2$ or less (Fig.\ref{fig:hrd_merge}). However, once the ratio got too small 
(e.g. $h_{\rm ov}$ being too large),  He core-burning took place in regions hotter than 
the blue-edge of the CIS.

Figure~\ref{fig:PWI_Mrgr} shows the periods of fundamental (filled blue circles) 
and first-overtone (blue asterisks) modes at the respective blue edges in the P-WI 
diagram showing again LMC observations in lighter grey. 
The size of the symbols reflects the mass of 
the model pulsators, ranging from $0.8$ to $1.2 \msol$. 
As mentioned before, sufficiently extended blue loops occur for $h_{\rm ov}$ choices 
within a narrow range only. Larger values of $h_{\rm ov}$ yield more luminous models, the 
resulting P-WI relation, as presented in Fig.~\ref{fig:PWI_Mrgr}, 
remained nonetheless insensitive to the particular choice of $h_{\rm ov}$. 

According to Fig.~\ref{fig:hrd_merge}, mergers apparently obey a unique $M-L$ relation 
in the region of the CIS; this translates into a tight P-WI relation with a well expressed 
mass segregation along the relation (Fig.~\ref{fig:PWI_Mrgr}). 
Lower-mass models were found at shorter periods and lower luminosities; the opposite applied 
to the more massive stars. The O1 P-WI relation is again well separated from the 
F mode; at constant period the O1 pulsators appear to be about $0.5$~mag brighter than 
an F-mode sibling (in other words, at a given brightness, the log of the period of an 
O1 pulsator is shorter by $-0.147$ compared with that of an F-mode pulsator; the period 
ratio of O1 to F mode is hence about $0.713$). The numerical values of the linear 
least-square fits to the `Case Merger' data are displayed in Tab.~\ref{tab:PWI-Fits}.
 
 
The He-core masses of $0.6 - 0.9 \msol$ at helium ignition of a merged star
can be thought to originate from two
stars with respective core masses on the FGB between about $0.3$ and $0.5 \msol$. 
An isolated $2 \msol$ star (without overshooting) develops a He core of about $0.23 \msol$ 
at the base of the FGB (at an age of $800$~Myrs) and it reaches about $0.42 \msol$ at 
He ignition (at about $860$~Myrs). A $4 \msol$ star develops a He core of about 
$0.5 \msol$ up to the onset of He burning at an age of about $140$~Myrs.
Therefore, the masses of the progenitors of our merger models should be $\la 4 \msol$ 
and the age of the merger products therefore older than about $200$~Myrs.  
 
\section{Discussion}
Remnants of close-binary interaction, be they donors, accretors, or even 
mergers have been shown to evolve into the CIS and to develop 
overstable radial pulsations with AC-compatible periods. 
We looked into the special case of the LMC because it hosts a respectable sample of ACs 
pulsating in either F or O1 modes. Furthermore, the LMC as a stellar system is commonly assigned 
a heavy-element abundance of about $Z=0.008$, with some spread in both directions. The scenario
of single-star evolution that takes stars through the CIS and produces He core-burning pulsators 
is challenged by stellar systems with abundances of heavy elements exceeding $Z\approx 0.001$ 
\citep{Fiorentino2012}. 

\begin{table}
	\centering
	\caption{Linear fit parameters (slope, intersection) of the 
	         $\log P$~--~Wesenheit-index relations of LMC ACs 
	         observed by OGLE and the blue edges of
			 the pulsating accretor components in the case of
			 Bc mass-transfer models ($Y=0.3, Z=0.001$).
			 Notice that no distance modulus is yet applied
			 to the intersections of the linear fits to the model data. 
			 }
	\label{tab:PWI-Fits}
	\begin{tabular}{lcc} 
		\hline
		Source      & F modes 			& O1 modes         \\
		\hline
		OGLE LMC    & (-2.97,15.59) 	& (-3.38, 16.03)   \\
		Case Bc-A   & (-3.16,-2.47) 	& (-3.26, -2.92)   \\
		Case Merger & (-3.50,-1.63)     & (-3.52, -2.16)   \\
		\hline
	\end{tabular}
\end{table}

In this paper we demonstrated that Case Bc MT generated accretor remnants 
with $Z\la 0.003$ and masses between $2.3$ and $3 \msol$ crossed the CIS with 
F and O1 periods compatible with observed ACs. The slopes of the P-WI relations 
are comparable to the observed one (cf. Fig.~\ref{fig:PWI_BcA}). 
Adopting, as always in this study, a distance modulus of 
18.5~mag to the LMC, the accretor P-WI relation is about $0.3$ to $0.4$~mag too bright, however. 

Considering donors as AC candidates, we found remnants with masses between $0.6$ and about
$1.1 \msol$ and with heavy element abundances as high as $Z=0.008$ that crossed the CIS with
F and O1 periods in the range of observed ACs in the LMC (cf. Fig.~\ref{fig:PWI_BcD}). 
The blue-edge data of the donors scattered considerably on the P-WI plane; not so much as to
be incompatible with observations, but too much to find a useful linear fit to the 
blue-edge data. Compared with the accretor models and with observed LMC ACs, the donor models 
were systematically too faint at their blue edges. We estimate the computed blue edges to be
subluminous by about 0.5 mag at a fixed period. The steepness of the computed blue-edge
P-WI data, nonetheless, agree decently with observations.    

Case Bc MT acts as an 
effective \emph{retardant} to the evolutionary appearance of 
accretor remnants and as an evolutionary \emph{catalyst} for donor remnants.
Hence, depending on which component of a binary system develops into an AC,
it can appear as either too young or too old compared with the dominating stellar
population of the harboring stellar system. 
Linking such stars to their binary membership, however, resolves the discrepancy.
Depending on the star-formation history of a stellar aggregate, the situation must be expected
to become rather complicated if a mixture of single-star and various close-binary scenarios 
contribute to the population of variable stars. 

We observed that the donor- and accretor-remnant ACs inherit their central properties 
from their precursors: Low-mass donor remnants started to burn helium nondegenerately 
in their centers whereas intermediate-mass  accretor remnants began helium burning off-center 
under degenerate conditions with one or a few flashes during which burning propagated to the 
stars' centers. Hence, in contrast to the single-star scenario for ACs,
binary remnants evolve into the CIS after either degenerate (i.e. He-flash)
or quiet non- or weakly-degenerate He ignition.

The possible case of 
MT resulting in the coalescence of the two binary components was adopted here 
as the \emph{merger} scenario contributing to the source of some ACs. The coarse modeling led
to AC-like pulsators for masses above about $0.8 \msol$ with sufficiently overgrown helium cores, 
exceeding about $0.6 \msol$. The mass of the helium cores was adjusted via an unusually large
overshooting parameter in the H-burning core of the precursors. Helium burning started quietly
in the merger models and blue-loops crossing the CIS were encountered for heavy-element abundances 
as high as $Z=0.008$. Hence, in a stellar system like the LMC merger remnants belonging even to a 
rather recent star generation could naturally contribute to its AC population.

The merger P-WI relations of the blue edges of the merger models are tight and the slopes
compare well to the LMC observations of OGLE project. As for the accretor and the donor cases, the
brightness of the relation at fixed period does not fit. In the merger case, 
the relation is subluminous, roughly at the level of the donor 
cases (compare Figs.~\ref{fig:PWI_Mrgr} and \ref{fig:PWI_BcD}).
In contrast to the accretor case, no obvious metallicity dependence was found of the blue-edge P-WI
relations. Static stellar models with unusually massive helium cores were constructed to
coarsely emulate the products of stellar mergers. 
The violent merging process, whose details probably depend on the masses and evolutionary 
states of the coalescing stars, was completely disregarded.
The system-specific differences during coalescence must be expected to affect the subsequent 
helium core-burning stage and to eventually inflict scatter on the P-WI relation.

Obvious Case A mass transfer was also looked into, and it was able to produce 
components that entered the CIS in the mass range of interest for ACs. The fine-tuning of the 
initial orbital separation was considerable, however. To find a way through semi-detached
binary evolution avoiding full contact, the orbital separation 
had to be finely tuned to within about $\pm 0.05 \rsol$. Therefore, we omitted such a path in this
first study because it seemed to be even more unlikely 
to be taken as a viable path to ACs compared to the binary scenarios we discussed here.

Details of the orbital separation of the binaries proved not to be crucial as 
long as the separation was chosen large enough so that the primary filled its Roche radius 
along the FGB with the convective envelope being deep to ensure robust case Bc MT. 


\citet{Karczmarek2016} pointed out recently that the influence of binary components, 
termed BEP (Binary Evolved Pulsators) in their paper, also in samples of classical pulsators 
should  not be undervalued. Stellar population synthesis simulations predict that a fair number of 
remnants of close binary interactions must be expected and that the fraction of such 
candidates increases towards higher luminosities in the CIS. Our study shows that binary 
components can be expected to be a part of the AC family of 
pulsators~--~complicating the quest to understand the origin of the 
group of variables; on the other hand, binary remnants open an alternative window to the star 
formation history of the hosting stellar system.

The Wesenheit~--~the essence~--~of the Wesenheit-index is its being
free of interstellar extinction, as long as the  ratio of extinction to color excess 
remains constant for the stars to which it is applied. It furthermore reduces 
the scatter of the data on the P-WI plane by diminishing the intrinsic 
width of the CIS because lines of constant period have a 
component along the extinction vector \citep[cf.][]{Madore1991}, which varies depending on the chosen photometric indices. Examining Fig.~\ref{fig:ACs_LMC} 
and comparing it with say Fig.~\ref{fig:PWI_LMCDCEP} or even better, inspecting
the observational data in the same panel \citep[Fig.~5 of][]{Soszynski2015},
reveals that AC F-mode pulsators in particular exhibit a substantial 
residual scatter which exceeds the one of the Classical Cepheids.
The effect of the finite width of the CIS should, however, be about the 
same for the two neighboring families of pulsators. Individual deviations 
of the ratio of reddening to color index of AC variables might, however, be a cause 
of the fan out of the P-WI relation; an definite answer requires
a star by star analysis. On the other hand, according to our exploratory computations, 
the presence of binary components in the AC population of the LMC would contribute 
to the blurring of the P-WI relation.

\section{Conclusions}
Although the results of this paper are far from exhaustive they can be considered 
a proof-of-concept that remnants of close-binary interactions
can make it through the CIS and pulsate radially with F- and O1-mode periods that compare 
well to observed ACs. Binary scenarios can hence complement the single-star
explanations of these variable stars with the additional benefit of higher stellar Z-abundances 
that still lead to CIS crossings.  

The mass range of AC candidates resulting from close-binary MT as computed here must 
be seen as lower bounds. Furthermore, we did not account for those low-mass stars whose 
loops extend into the CIS but do not extend far enough to cross the respective blue edges, and 
we neglected non-conservative mass-transfer. 

If binary components are also to be admitted in the AC family then a much less well-defined 
mass-luminosity relation must be expected to be encountered. Specifically, the masses 
of AC variables would embrace a broader range than anticipated hitherto, 
when invoking single-star evolution only \citep[e.g.][]{Fiorentino2012}. 

The binary orbits of the post~--~Bc MT models were seen to grow to the order of $300 - 600$ 
days by the time the remnants reached the instability strip; 
accordingly, the associated orbital speeds of ACs are as low as a few kilometers per second. 
The observable binary indicators are therefore weak enough so that
they might have been either technically inaccessible or having been overlooked so far.
One prediction of our scenario is therefore that those ACs that descend from close binary 
interactions via a case Bc mass transfer show, caused by their wide binary orbit, 
radial velocity variations on the time scale of several hundred days with amplitudes 
of a few kilometers per second. 
If the \emph{Gaia }mission can contribute to a clarify the situation remains to be seen; 
after all, a proper phasing of pointings would be necessary to get a decent radial-velocity 
coverage that allows to disentangle pulsations and binary orbit.

In the case of donors turned to ACs, we predict that cases of ACs with helium-enriched envelopes 
should be observed; such cases might serve as spectroscopic tracers of their evolutionary past.     

The modest CPU power which was available for this proof-of-concept study required simplifications
on various levels of modeling. Systematic surveys of a much larger binary parameter space are,
though, readily accessible to more capable computer systems. 
The data from such approaches are necessary to solidify the picture advocated here.

The persisting, complete absence of double-mode ACs in the observations 
of all stellar systems and the Galactic field remains intriguing. 
Linear modeling shows no signs of structural anomalies that 
hint at unusual excitation properties of either F or O1 modes; to the best of the
authors' knowledge also the nonlinear modeling has been of no help so far in this respect.

\section*{Acknowledgements}
This research has made use of NASA's Astrophysics Data System Bibliographic Services. 
The computations were made possible by the generous open-source philosophy of the 
MESA and GYRE code developers. 

\bibliographystyle{mnras}
\bibliography{ACBib}  

\appendix
\section{Ansatz-Validation with LMC Classical Cepheids}
\label{sec:validation}

\begin{figure}
	\includegraphics[width=\columnwidth]{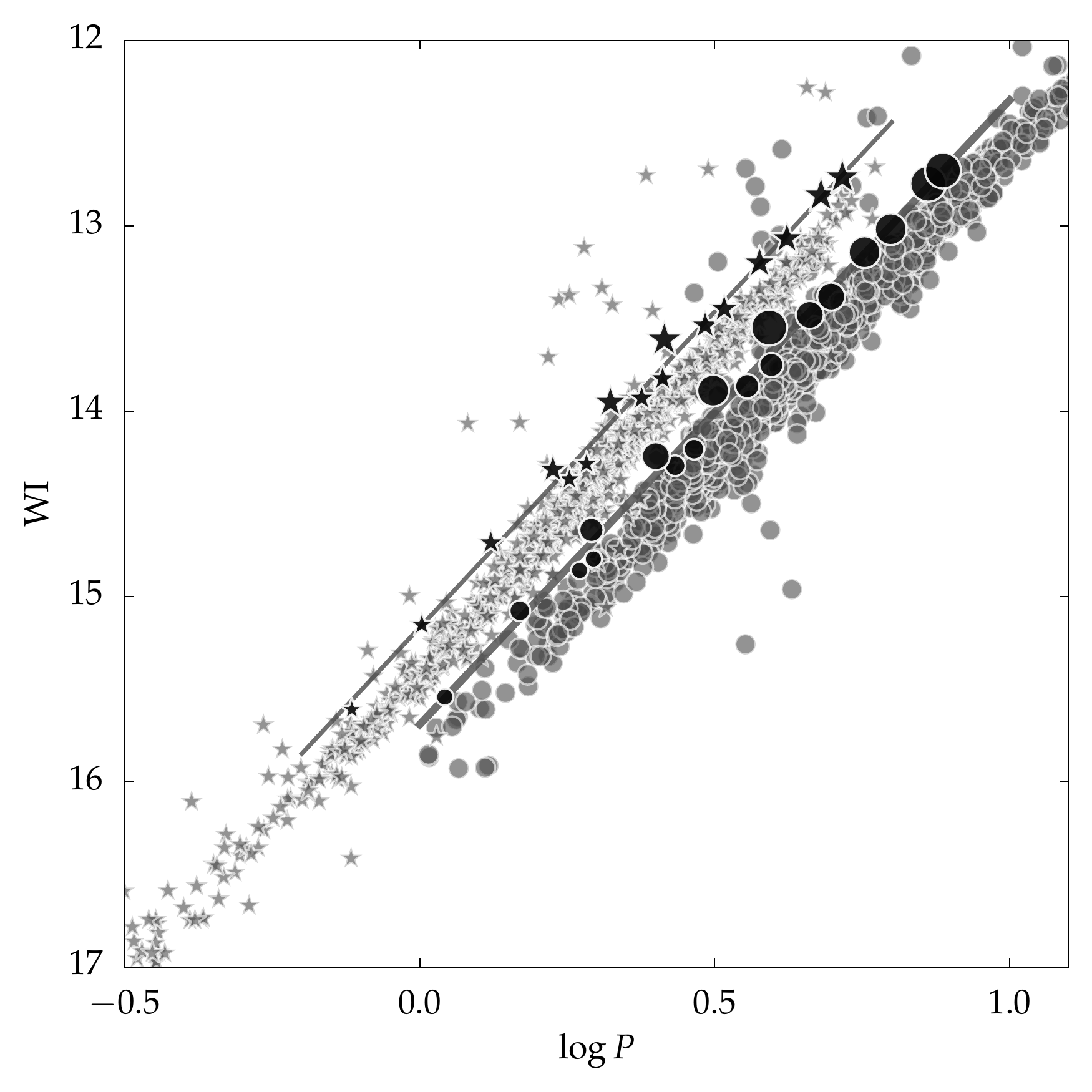}
    \caption{The P-WI diagram for LMC Classical Cepheids according to 
             OGLE observations; filled grey circles show
    	     F-mode, and grey asterisks O1-mode pulsators. 
    	     GYRE results for the blue edges of the instability regions 
    	     of F- and O1-modes are superimposed as black circles and asterisks, 
    	     respectively. The model results are connected with a linear 
    	     least-square fit.}
    \label{fig:PWI_LMCDCEP}
\end{figure}

To inspect the performance of the simplified pulsation treatment 
adhered to in this paper a few intermediate-mass star models were 
evolved into their Cepheid phase. 
Observations from the OGLE project give access to a rich 
population of Classical Cepheids in the LMC \citep{Soszynski2008a}, 
which served as fiducial data set. 
Model data at the blue edges of the F- and O1-mode instability regions on 
the P-WI plane were computed to be compared with the OGLE observations in 
Fig.~\ref{fig:PWI_LMCDCEP}.

The model data for Fig.~\ref{fig:PWI_LMCDCEP} were computed on evolutionary models 
with $Z=0.008$, Schwarzschild criterion to treat convection adopting 
$\alpha_{\mathrm{MLT}}=1.5$, as suggested in \citet{Baraffe1998}. 
The F- and O1-mode blue-edge data points plotted in the figure result 
from first to third crossing of the CIS of model stars with masses 
between $4$ and $6.5 \msol$; the marker sizes scale with stellar mass. 

Even though the perturbation of the convective flux in the 
pulsation equations was neglected, Fig.~\ref{fig:PWI_LMCDCEP} proves that location and 
slopes of F- and O1-mode blue edges agree very well with the observed 
LMC Classical Cepheids having periods shorter than
roughly $10$~days. The slopes of the linear fits to the blue-edge computations
and the slopes of the observed F- and O1-pulsators' 
ridge-lines differ by less than $11 \%$. 
Visual inspection of Fig.~\ref{fig:PWI_LMCDCEP} shows that the
computed positions of the blue-edges also perform well as borders to
observational data when adopting a distance modulus to the LMC of $18.5$ magnitudes again.  

The LMC ACs pulsate with periods typically shorter than $4$~days. Hence, 
they are hotter than most of the Classical Cepheids considered here and 
they have shallower convection zones, so that the effect of convection on structure
and pulsations is further diminished. We are therefore positive that the 
frozen-in convective flux treatment in the pulsation equations is good 
enough for this study of effect of binary components on the AC 
variable-star population. 
 
\bsp	
\label{lastpage}
\end{document}